\documentclass{article}

\usepackage{amssymb}

\newbox\tempa
\newbox\tempb
\newdimen\tempc
\def\mud#1{\hfil $\displaystyle{\mathstrut #1}$\hfil}
\def\rig#1{\hfil $\displaystyle{#1}$}
\def\irulehelp#1#2#3{\setbox\tempa=\hbox{$\displaystyle{\mathstrut #2}$}%
                        \setbox\tempb=\vbox{\halign{##\cr
        \mud{#1}\cr
        \noalign{\vskip\the\lineskip}%
        \noalign{\hrule height 0pt}%
        \rig{\vbox to 0pt{\vss\hbox to 0pt{${\; #3}$\hss}\vss}}\cr
        \noalign{\hrule}%
        \noalign{\vskip\the\lineskip}%
        \mud{\copy\tempa}\cr}}%
                      \tempc=\wd\tempb
                      \advance\tempc by \wd\tempa
                      \divide\tempc by 2 }
\def\irule#1#2#3{{\irulehelp{#1}{#2}{#3}%
                     \hbox to \wd\tempa{\hss \box\tempb \hss}}}

\def\imp{\rightarrow}
\def\sequent{\vdash}

\newtheorem{definition}{Definition}[section]

\newtheorem{proposition}{Proposition}[section]

\newenvironment{example}{\noindent{\em Example.}}{}
\newenvironment{remark}{\noindent{\em Remark.}}{}

\newcommand{\proof}[1]{{{\em Proof.} #1 $\Box$}}

\begin{document}
\title{Eigenvariables, bracketing and the decidability of
positive minimal predicate logic}
\author{Gilles Dowek\thanks{\'Ecole polytechnique and INRIA,
                  LIX, \'Ecole polytechnique,
                  91128 Palaiseau Cedex, France.
             {\tt Gilles.Dowek@polytechnique.fr}}
     ~and Ying Jiang\thanks{Institute of Software, Chinese Academy of
Sciences, P.O. Box 100080 Beijing, China. {\tt jy@ios.ac.cn}} }
\date{}
\maketitle

\thispagestyle{empty}

\begin{abstract}
We give a new proof of a theorem of Mints that the positive
fragment of minimal predicate logic is decidable. The idea of the
proof is to replace the eigenvariable condition of sequent
calculus by an appropriate scoping mechanism. The algorithm given
by this proof seems to be more practical than that given by the
original proof. A naive implementation is given at the end of the
paper. Another contribution is to show that this result extends to
a large class of theories, including simple type theory
(higher-order logic) and second-order propositional logic. We
obtain this way a new proof of the decidability of the
inhabitation problem for positive types in system F.

\end{abstract}

\def\lastname{Dowek and Jiang}

\section*{Introduction}

In classical propositional logic, the rules of sequent calculus
can be chosen in order to commute with contraction and thus a
sequent has a derivation if and only if it has a cut-free
contraction-free derivation. The search space for cut-free
contraction-free derivations is finite and hence classical
propositional logic is decidable.

In minimal propositional logic, the left rule of the implication
does not commute with contraction anymore and thus to remain
complete when searching for a derivation, we have to duplicate an
implication occurring in the left part of a sequent before we
decompose it. For instance, to prove the formula $((((P \imp Q) \imp
P) \imp P) \imp Q) \imp Q$ it is necessary to use the formula $(((P
\imp Q) \imp P) \imp P) \imp Q$ twice (see Definition
\ref{sequentcalculus} below for the sequent calculus used in this
paper).
$$\irule{\irule{\irule{\irule{\irule{\irule{\irule{\irule{}
                                                         {(((P \imp Q) \imp P) \imp P) \imp Q, (P \imp Q) \imp P, P \sequent P}
                                                         {\mbox{\mbox{$L\imp$}}}
                                                  }
                                                  {(((P \imp Q) \imp P) \imp P) \imp Q, (P \imp Q) \imp P, P \sequent ((P \imp Q) \imp P) \imp P}
                                                  {\mbox{$R\imp$}}
                                           }
                                           {(((P \imp Q) \imp P) \imp P) \imp Q, (P \imp Q) \imp P, P \sequent Q}
                                           {\mbox{\mbox{$L\imp$}}}
                                    }
                                    {(((P \imp Q) \imp P) \imp P) \imp Q, (P \imp Q) \imp P \sequent P \imp Q}
                                    {\mbox{$R\imp$}}
                             }
                             {(((P \imp Q) \imp P) \imp P) \imp Q, (P \imp Q) \imp P \sequent P}
                             {\mbox{\mbox{$L\imp$}}}
                      }
                      {(((P \imp Q) \imp P) \imp P) \imp Q \sequent ((P \imp Q) \imp P) \imp P}
                      {\mbox{$R\imp$}}
               }
               {(((P \imp Q) \imp P) \imp P) \imp Q \sequent Q}
               {\mbox{$L\imp$}}
         }
         {\sequent ((((P \imp Q) \imp P) \imp P) \imp Q) \imp Q}
         {\mbox{$R\imp$}}$$
This derivation yields the long normal proof-term $$\lambda
\alpha^{(((P \imp Q) \imp P) \imp P) \imp Q}~(\alpha~\lambda \beta^{(P
\imp Q) \imp P}~(\beta~\lambda \gamma^{P}~(\alpha~\lambda \beta'^{(P
\imp Q) \imp P}~\gamma)))$$ where the variable $\alpha$ is used
twice.

Thus, the decidability of minimal propositional logic is not as
obvious as that of classical propositional logic, and to design a
decision algorithm for minimal propositional logic or for the
inhabitation problem in simply typed lambda-calculus, we need
either to restrict to non redundant proofs, as, for instance, in 
\cite{Kleene}, or to specialize sequent
calculus to avoid this left rule of the implication, as for
instance, in \cite{Dyckhoff}.

When we extend classical propositional logic by allowing positive
quantifiers ({\em i.e.} universal quantifiers at positive occurrences
and existential quantifiers at negative occurrences), we need to
introduce two more rules in sequent calculus: the right rule of
the universal quantifier and the left rule of the existential
quantifier. These rules also commute with contraction, hence, the
positive fragment of classical predicate logic is decidable too.
Another way to put the argument is that, in classical logic, any
formula with positive quantifiers can be transformed into a prenex
universal formula, hence provability in the positive fragment can
be reduced to provability in the propositional fragment.

If we have negative quantifiers also, we need to introduce two
more rules: the left rule of the universal quantifier and the
right rule of the existential quantifier. These rules do not
commute with contraction and the decidability result does not
extend.  The fact that, in classical predicate logic, contraction
needs to be applied only below these two rules can be seen as a
formulation of Herbrand's theorem.

When we extend minimal propositional logic with positive
quantifiers, the situation is again more complicated. For instance
in the derivation
$$\irule{\irule{\irule{\irule{\irule{\irule{\irule{\irule{}{A,
(Q \imp R) \imp Q, P(x), Q, (Q \imp R) \imp Q, P(x') \sequent
Q}{\mbox{$L\imp$}}}
                                           {A, (Q \imp R)
\imp Q, P(x), Q \sequent \forall x~(((Q \imp R) \imp Q) \imp P(x)
\imp Q)}
                                           {\mbox{$R\imp$, $R\forall$}}
                                    }
                                    {A, (Q \imp R) \imp Q, P(x), Q \sequent R}
                                    {\mbox{$L\imp$}}
                             }
                             {A, (Q \imp R) \imp Q, P(x) \sequent Q \imp R}
                             {\mbox{$R\imp$}}
                      }
                      {A, (Q \imp R) \imp Q, P(x) \sequent Q}
                      {\mbox{$L\imp$}}
               }
               {A
               \sequent \forall x~(((Q \imp R) \imp Q) \imp P(x) \imp Q)}
               {\mbox{$R\imp$, $R\forall$}}
        }
        {A \sequent R}
        {\mbox{$L\imp$}}}
        {\sequent A \imp R}
        {\mbox{$R\imp$}}$$
where $A$ is the formula $(\forall x~(((Q \imp R)\imp Q) \imp P(x)
\imp Q)) \imp R$, we need to rename the variable $x$ into $x'$ when
applying the right rule of the universal quantifier for the second
time. The proof-term associated to this derivation is
$$\lambda \alpha^A (\alpha~\lambda x \lambda \beta^{(Q \imp R) \imp Q}\lambda \gamma^{P(x)}~(\beta~\lambda \delta^{Q}~(
\alpha~\lambda x' \lambda \beta'^{(Q \imp R) \imp Q} \lambda
\gamma'^{P(x')}~\delta)))$$ Thus, not only $\alpha$ occurs
twice in this term, but also each occurrence yields a different
bound variable: $x$ and $x'$.

Hence the formul\ae\  that may occur in the derivations are not in
a finite space anymore and, even when restricted to non redundant
proofs, proof search 
may fail to terminate. For instance, searching for a derivation of
the formula
$$((\forall x~(P(x) \imp Q)) \imp Q) \imp Q$$
we develop the following attempt where $A$ is the formula
$(\forall x~(P(x) \imp Q)) \imp Q$.
$$\irule{
\irule{\irule{\irule{\irule{\irule{\irule{\irule{...}
                                                  {A, P(x), P(x'), P(x'') \sequent Q}
                                                  {}
                                           }
                                           {A, P(x), P(x') \sequent \forall x~(P(x) \imp Q)}
                                           {\mbox{$R\imp$, $R\forall$}}
                                    }
                                    {A, P(x), P(x') \sequent Q}
                                    {\mbox{$L\imp$}}
                               }
                                    {A, P(x) \sequent \forall x~(P(x) \imp Q)}
                                    {\mbox{$R\imp$, $R\forall$}}
                             }
                             {A, P(x) \sequent Q}
                             {\mbox{$L\imp$}}
               }
               {A \sequent \forall x~(P(x) \imp Q)}
               {\mbox{$R\imp$, $R\forall$}}
        }
        {A \sequent Q}
        {\mbox{$L\imp$}}
} {\sequent A \imp Q} {\mbox{$R\imp$}}$$ In this attempt,
we accumulate formul\ae\  $P(x)$, $P(x')$, $P(x'')$, ... and naive
restriction to non redundant proofs fails to prune this branch.

Notice that, in minimal predicate logic, the provability of a
formula is not equivalent to the provability of its prenex form,
so we cannot reduce provability in the positive fragment to
provability in the propositional fragment by putting the formula
to be proved in prenex form. For instance, the formula
\[
\forall x~(((\forall y \forall z~((R(y,x) \imp P(z)) \imp (R(y,z)
\imp P(z)))) \imp P(x))\imp P(x))
\]
where $R(y,x) = P(y) \imp P(x)$ and $R(y,z) = P(y) \imp P(z)$, is
not derivable, although its prenex form
$$\forall x \forall y \forall
z~((((R(y,x) \imp P(z)) \imp (R(y,z) \imp P(z))) \imp P(x)) \imp
P(x))$$ is.

Mints \cite{Mints} proves that, in the positive fragment of
intuitionistic predicate logic, a provable formula always has a
derivation with less than $n$ variables, where $n$ is a bound
computed as a function of the formula.  This way, the search space
can be restricted to be finite and hence the positive fragment of
intuitionistic predicate calculus is proved to be decidable.

We know that, in logic, variable names are irrelevant and that
replacing named variables by another scoping mechanism, such as de
Bruijn indices \cite{DeBruijn}, simplifies formalisms very often.

The goal of this paper is to replace the eigenvariable condition
of the sequent calculus, that forces to rename bound variables and
to invent new variable names, by an alternative scoping mechanism.
We obtain this way an alternative decision algorithm for the
positive fragment of minimal predicate logic, where the search
space is restricted just by restricting to non redundant proofs,
like in the
propositional case. A naive implementation of this algorithm is
given at the end of the paper.

For sake of simplicity, we consider only minimal logic in this
paper, but the method developed should extend smoothly to full
intuitionistic logic. However, we leave this extension for future
work.

Finally, we show that our decidability result extends to simple
type theory (higher-order logic) and to system F. We obtain this
way a new algorithm testing inhabitation of positive types in
system F \cite{Jiang}. 

Notice that the encoding of traditional data types in system F (such
as the empty data type $\forall X~X$, booleans $\forall X~(X \imp (X
\imp X))$ and natural numbers $\forall X~(X \imp ((X \imp X) \imp
X))$) are positive types. Thus this decidability result raises the
question of the possibility to consider all positive types as extended
data types.

A preliminary version of this paper appeared in \cite{DowekJiang}.
The system presented in this paper is simpler than that of
\cite{DowekJiang} because we deal directly with variables instead
of using the technical notion of {\em level} previously used.

\section{Positive formul\ae}

In {\em minimal predicate logic}, the syntax of {\em terms} and
{\em formul\ae} is given by
$$t = x~|~f(t, ..., t)$$
$$A = P(t, ..., t)~|~(A \imp A)~|~\forall x~A$$

Superfluous parentheses are omitted as usual.
Free and bound occurrences of variables in a formula are defined as usual. 

Formally, a formula $A$ is a tree, whose nodes are labeled with either
an atomic formula $P(t_{1}, ..., t_{n})$ or the symbol $\imp$ or else
the quantifier $\forall$ and a variable.

To each position in such a tree, we associate a formula. These
formul\ae\  are called {\em the pieces} of $A$. For instance the
pieces of the formula $\forall x~(P(x) \imp Q)$ are $\forall
x~(P(x) \imp Q)$, $P(x) \imp Q$, $P(x)$ and $Q$. Notice that this
notion of piece is different from the usual notion of sub-formula,
as we cannot substitute for the variables in pieces.

A {\em LJ$^{+}$-context} is a finite multiset of formul\ae.
A {\em LJ$^{+}$-sequent} $\Gamma \sequent A$ is a pair formed
by a context $\Gamma$ and a formula $A$.

\begin{definition}[Free and bound variables of a context]
{\em Free} and {\em bound} variables of a context are defined by
\begin{itemize}
\item  $FV(\{A_{1}, ..., A_{n}\}) = FV(A_{1}) \cup ... \cup
FV(A_{n})$,
\item  $BV(\{A_{1}, ..., A_{n}\}) = BV(A_{1}) \cup ... \cup
BV(A_{n})$.
\end{itemize}
\end{definition}

A formula in minimal predicate logic is positive if all its
universal quantifier occurrences are positive. More precisely, the
set of positive and negative formul\ae\  are defined by induction
as follows.

\begin{definition}{\bf (Positive and negative formul\ae\ and sequents)}
\begin{itemize}
\item An atomic formula is {\em positive} and {\em negative}, 
\item a formula of the form $A \imp B$ is {\em positive} 
(resp. {\em negative}) if $A$ is {\em negative} (resp. {\em positive}) 
and $B$ is {\em positive} (resp. {\em negative}),

\item a formula of the form $\forall x~A$ is {\em positive} if $A$ is
{\em positive},
\item 
a sequent $A_{1}, ..., A_{n} \sequent B$ is {\em positive} if
$A_{1}$, ..., $A_{n}$ are {\em negative} and $B$ is {\em positive}.
\end{itemize}
\end{definition}

Notice that a formula of the form $\forall x~A$ is never negative.

\begin{proposition}
A negative formula has the form $A_{1} \imp ... \imp A_{n} \imp P$
where $P$ is an atomic formula and $A_{1}, ..., A_{n}$ are
positive formul\ae.
\end{proposition}

We use a cut-free sequent calculus for positive sequents  in
minimal predicate logic. Instead of the usual axiom rule
$$\irule{}
        {\Gamma, A \sequent A}
        {}$$
and left rule for implication
$$\irule{\Gamma, A \imp B \sequent A~~~\Gamma, A \imp B, B
\sequent C}
        {\Gamma, A \imp B \sequent C}
        {}$$
we take a more restricted rule, in the style of Howard,
$$\irule{\Gamma, A_{1} \imp ... \imp  A_{n} \imp
P \sequent A_{1}~~...~~\Gamma, A_{1} \imp ... \imp A_{n} \imp P
\sequent A_{n}}
        {\Gamma, A_{1} \imp ... \imp  A_{n} \imp
P \sequent P}
        {}$$
where $P$ is an atomic formula. This way, derivations can be
directly translated to long normal proofs in natural deduction,
and the formula $A_{1} \imp ... \imp  A_{n} \imp P$ is the type of
the head variable of the associated proof-term.

The equivalence of this system with that having the usual axiom and
$L\imp$ rules is straightforward.

In this sequent calculus, formul\ae\  are, as usual, identified
modulo $\alpha$-equivalence.

\begin{definition}{\bf (LJ$^{+}$, A sequent calculus for positive sequents)}
\label{sequentcalculus}

$$\irule{\Gamma, A_{1} \imp ... \imp A_{n} \imp P  \sequent
A_{1}~~~...~~~\Gamma, A_{1} \imp ... \imp A_{n} \imp P \sequent
A_{n}}
        {\Gamma, A_{1} \imp ... \imp A_{n} \imp P
\sequent P}
        {\mbox{\em $L\imp$}}$$
if $P$ is atomic.
$$\irule{\Gamma \sequent A}
        {\Gamma \sequent \forall x~A}{\mbox{\em $R\forall$}}$$
if $x$ is not free in $\Gamma$.
$$\irule{\Gamma, A \sequent B}
        {\Gamma \sequent A \imp B}{\mbox{\em $R\imp$}}$$
\end{definition}

\section{LJB : a sequent calculus with brackets}
\label{sectionLJB}

In LJ$^{+}$, when we have a sequent of the form $\Gamma
\sequent \forall x~A$, we may need to rename the variable $x$
with a variable $x'$ that is free neither in $\Gamma$ nor
in $A$ in order to apply the $R\forall$ rule. We introduce now
another sequent calculus, where, instead of renaming the variable
$x$, we bind it in the context $\Gamma$ with brackets and obtain
the sequent $[\Gamma]_{x} \sequent A$.

In fact, we will bind in $\Gamma$, not only the variable $x$, but
also all the variables bound in $A$. Although binding $x$
only and binding all the variables bound in $\forall x~A$ both yield a
sound and complete system, this second choice allows to prove
termination of proof search.

\begin{definition}{\bf (LJB-contexts and items)}
{\em LJB-contexts} and {\em items} are mutually inductively
defined as follows.
\begin{itemize}
\item A {\em LJB-context} $\Gamma$ is a finite multiset of items
$\{I_{1}, ..., I_{n}\}$,

\item an {\em item} $I$ is either a formula or an expression of
the form $[\Gamma]_{V}$ where $V$ is a set of variables and $\Gamma$ a
context.
\end{itemize}
\end{definition}

In the item $[\Gamma]_{V}$ the variables of $V$ are bound by the
symbol $[~]$.

\begin{definition}[Free and bound variables of a LJB-context and of an item]
The set of {\em free variables} of a LJB-context is defined by
\begin{itemize}
\item $FV(\{I_{1}, ..., I_{n}\}) =  FV(I_{1}) \cup ... \cup
 FV(I_{n})$,
\end{itemize}
and the set of {\em free variables} of an item by
\begin{itemize}
\item $FV(A) =  FV(A)$, \item $FV([\Gamma]_{V}) = FV(\Gamma)
\setminus V$.
\end{itemize}

The set of {\em bound variables} of a LJB-context is defined by
\begin{itemize}
\item $BV(\{I_{1}, ..., I_{n}\}) =  BV(I_{1}) \cup ... \cup
 BV(I_{n})$,
\end{itemize}
and the set of {\em bound variables} of an item by
\begin{itemize}
\item $BV(A) =  BV(A)$, \item $BV([\Gamma]_{V}) = BV(\Gamma) \cup
V$.
\end{itemize}
\end{definition}

A {\em LJB-sequent} $\Gamma \sequent A$ is a pair formed by a
LJB-context $\Gamma$ and a formula $A$.

The system LJB is formed by two sets of rules: the usual deduction
rules and additional transformation rules. The transformation
rules deal with bracket manipulation. They form a terminating rewrite
system. The first transformation rule allows to 
replace an item of the form $[I, \Gamma]_{V}$ by the two items $I$ and
$[\Gamma]_{V}$ provided no free variable of $I$ is in $V$. The second
rule allows to remove trivial items. 
The third rule to replace two identical items by one.

\begin{definition}{\bf (Cleaning LJB-contexts)}
\label{cleaning} We consider the following rules simplifying
LJB-contexts, where $I$ is a item and $\Gamma$ a LJB-context.

\[\begin{array}{ll}
[I, \Gamma]_{V} \longrightarrow I, [\Gamma]_{V}, & \mbox{if
\(FV(I) \cap V = \emptyset\)}\\
\left[\ \right]_{V} \longrightarrow \emptyset &\\
I I \longrightarrow I&
\end{array}
\]
\end{definition}

As usual, these rules may be applied anywhere in a LJB-context.

\begin{proposition}[Termination]
The rewrite system of Definition \ref{cleaning} terminates.
\end{proposition}

\proof{We check that the following interpretation decreases
\begin{itemize}
\item $|A| = 1$, if $A$ is a formula, \item $|[\Gamma]_{V}| = 1+ 2
|\Gamma|$, \item $|I_{1}, ..., I_{n}| = |I_{1}| + ... + |I_{n}|$.
\end{itemize}}

The confluence of this system is more tricky. Indeed, although it is
quite simple, this rewrite system is defined modulo associativity and
commutativity (as contexts are multisets), it contains a binding
symbol and it is non linear. Thus, instead of proving confluence,
we shall fix an arbitrary strategy and define the normal form $\Gamma
\downarrow$ of a context $\Gamma$ as the normal form relative to this
strategy.

We now turn to the deduction rules. These rules apply only to
normal LJB-sequents where in each formula the bound variables are
distinct and distinct from all the free variables. It is easy to
check that these properties are preserved by the rules. Notice
also that in LJB we deal with formul\ae, not formul\ae\  modulo
$\alpha$-equivalence.

\begin{definition}{\bf (LJB, A sequent calculus with brackets)}
\label{bracketedsequentcalculus}

$$\irule{\Gamma' \sequent A_{1} ~~~ ...  ~~~\Gamma' \sequent A_{n}}
        {\Gamma \sequent P} {\mbox{\em $L\imp$}}$$
where
$$\Gamma = \Gamma_{1}, [\Gamma_{2}, [... \Gamma_{i-1},
[\Gamma_{i}, A_{1} \imp ... \imp A_{n} \imp P]_{V_{i-1}}
...]_{V_{2}}]_{V_{1}}$$
$$\Gamma' = ([...[[\Gamma_{1}]_{V_{1}}, \Gamma_{2}]_{V_{2}}, ...
\Gamma_{i-1}]_{V_{i-1}}, \Gamma_{i}, A_{1} \imp ... \imp A_{n} \imp
P) {\downarrow}$$ $P$ is atomic and has no free variable in
$V_{1} \cup V_{2} \cup ... \cup V_{i-1}$.
$$\irule{[\Gamma]_{V} {\downarrow} \sequent A}
        {\Gamma \sequent \forall x~A}
        {\mbox{\em $R\forall$}}$$
        where $V$ is the set of all variables bound in $\forall x~A$
$$\irule{(\Gamma, A) {\downarrow} \sequent B}{\Gamma \sequent A \imp B}{\mbox{\em $R\imp$}}$$
\end{definition}

In the $L\imp$ rule, brackets are moved from some items of
the LJB-context to others, bringing the formula $A_{1} \imp ... \imp
A_{n} \imp P$ inside brackets to the surface, so that it can be
used. For instance the LJB-sequent $Q(x), [Q(x) \imp P]_{x}
\sequent P$ is transformed (bottom-up) into $[Q(x)]_{x}, Q(x)
\imp P \sequent Q(x)$. The crucial point is that the two
occurrences of $x$ in $Q(x)$ and $Q(x) \imp P$ that are separated
in the first LJB-sequent remain separated. This idea is made
precise in Proposition \ref{invariant2}.

When $i = 1$ the $L\imp$ rule degenerates to
$$\irule{\Gamma_{1}, A_{1} \imp ...
\imp A_{n} \imp P \sequent A_{1} ~~~ ... ~~~\Gamma_{1}, A_{1} \imp
... \imp A_{n} \imp P \sequent A_{n}}
        {\Gamma_{1}, A_{1} \imp ... \imp A_{n} \imp
P \sequent P} {\mbox{\em $L\imp$}}$$ with $P$ atomic.

\begin{example}
Let us try again to prove the formula
$$((\forall x~(P(x) \imp Q)) \imp Q) \imp
Q$$
We obtain the following attempt
$$\irule{
\irule{\irule{\irule{\irule{\irule{\irule{\irule{...}
                                                {A, [P(x)]_{x}, P(x) \sequent Q}
                                                {}
                                         }
                                           {A, [P(x)]_{x}, P(x) \sequent \forall x~(P(x) \imp Q)}
                                           {\mbox{$R\imp$, $R\forall$}}
                                   }
                                   {A, [P(x)]_{x}, P(x) \sequent Q}
                                   {\mbox{$L\imp$}}
                           }
                           {A, P(x) \sequent \forall x~(P(x) \imp Q)}
                           {\mbox{$R\imp$, $R\forall$}}
                   }
                   {A, P(x) \sequent Q}
                   {\mbox{$L\imp$}}
            }
            {A \sequent \forall x~(P(x) \imp Q)}
            {\mbox{$R\imp$, $R\forall$}}
        }
        {A \sequent Q}
        {\mbox{$L\imp$}}
} {\sequent A \imp Q} {\mbox{$R\imp$}}$$ where $A$ is the
formula $(\forall x~(P(x) \imp Q)) \imp Q$.

Now, instead of accumulating formul\ae\  $P(x)$, $P(x')$,
$P(x'')$, ... we accumulate items $[P(x)]_{x}$, that collapse by
context cleaning. Thus, the LJB-sequent $A, [P(x)]_{x}, P(x)
\sequent Q$ is repeated and restricting to non redundant proofs prunes
this branch.
\end{example}

\begin{example}
Let us try now a slightly more involved example
$$((\forall x~((P(x) \imp Q) \imp Q)) \imp Q) \imp Q$$ 
We obtain the following attempt
$$\irule{\irule{\irule{\irule{...}
                             {A, P(x) \imp Q \sequent Q}
                             {}
                      }
                      {A \sequent \forall x~((P(x) \imp Q) \imp Q)}
                      {\mbox{$R\imp$, $R\forall$}}
               }
               {A \sequent Q}
               {\mbox{$L\imp$}}
        }
        {\sequent A \imp Q}
        {\mbox{$R\imp$}}$$
where $A$ is the formula $(\forall x~((P(x) \imp Q) \imp Q)) \imp Q$.

At this point, we have two possibilities. Either we use the
$L\imp$ rule with the formula $P(x) \imp Q$ and we have to
prove the LJB-sequent $A, P(x) \imp Q \sequent P(x)$ to which no
rule applies, or we use this same rule with the formula $A$ and we
have to prove the LJB-sequent $A, P(x) \imp Q \sequent \forall
x~((P(x) \imp Q) \imp Q)$, in this case the search continues as
follows
$$\irule{\irule{\irule{\irule{\irule{\irule{...}
                                           {A, [P(x) \imp Q]_x, P(x) \imp Q \sequent Q}
                                           {}
                                     }
                                     {A, P(x) \imp Q \sequent \forall x~((P(x) \imp Q) \imp Q)}
                                     {\mbox{$R\imp$, $R\forall$}}
                             }
                             {A, P(x) \imp Q \sequent Q}
                             {\mbox{$L\imp$}}
                     }
                      {A \sequent \forall x~((P(x) \imp Q) \imp Q)}
                      {\mbox{$R\imp$, $R\forall$}}
               }
               {A \sequent Q}
               {\mbox{$L\imp$}}
        }
        {\sequent A \imp Q}
        {\mbox{$R\imp$}}$$
At this point, we have three possibilities. The first is to use
the $L\imp$ rule with the formula $P(x) \imp Q$ and we have
to prove the LJB-sequent $$A, [P(x) \imp Q]_x, P(x) \imp Q
\sequent P(x)$$ to which no rule applies. The second is to use
this same rule with the formula $P(x) \imp Q$ inside the brackets.
In this case we have to move the brackets and we obtain the
LJB-sequent
$$A, P(x) \imp Q, [P(x) \imp Q]_x \sequent P(x)$$ to
which no rule applies. The third is to use this same rule with the
formula $A$ and we have to prove the LJB-sequent
$$A, [P(x) \imp Q]_x, P(x) \imp Q \sequent \forall
x~((P(x) \imp Q) \imp Q)$$ in this case the search continues as
follows
$$\irule{\irule{\irule{\irule{\irule{\irule{\irule{\irule{...}
                                                         {A, [P(x) \imp Q]_x, (P(x) \imp Q) \sequent Q}
                                                         {}
                                                  }
                                                  {A, [P(x) \imp Q]_x, P(x) \imp Q
\sequent \forall x~((P(x) \imp Q) \imp Q)}
                                                  {\mbox{$R\imp$, $R\forall$}}
                                            }
                                            {A, [P(x) \imp Q]_x, P(x) \imp Q \sequent Q}
                                            {\mbox{$L\imp$}}
                                     }
                                     {A, P(x) \imp Q \sequent \forall x~((P(x) \imp Q) \imp Q)}
                                     {\mbox{$R\imp$, $R\forall$}}
                              }
                              {A, P(x) \imp Q \sequent Q}
                              {\mbox{$L\imp$}}
                      }
                      {A \sequent \forall x~((P(x) \imp Q) \imp Q)}
                      {\mbox{$R\imp$, $R\forall$}}
               }
               {A \sequent Q}
               {\mbox{$L\imp$}}
        }
        {\sequent A \imp Q}
        {\mbox{$R\imp$}}$$
and the branch is pruned because the sequent $A, [P(x) \imp Q]_x,
(P(x) \imp Q) \sequent Q$ appears previously.
\end{example}

\section{Equivalence}

In this section, we want to prove the equivalence of the systems
LJ$^{+}$ and LJB, {\em i.e.} if $A$ is a formula, then the sequent
$\sequent A$ is provable in LJB if and only if it is provable
in LJ$^{+}$.

Several methods can be used to prove this equivalence. Some are
based on model constructions and others are based on proof
transformations. In proof transformation based methods, we first
have to define a translation of LJB-sequents to LJ$^{+}$-sequents.
Again there are several possibilities. One of them is to introduce
existential quantifiers to replace the brackets and translate, for
instance, the sequent $$[P(x) \imp P(y)]_{x,y}, [P(x)]_{x}
\sequent P(z)$$ to $$\exists x \exists y~(P(x) \imp P(y)),
\exists x~P(x) \sequent P(z)$$

We have chosen another translation by introducing a distinct free
variable for each variable bound by brackets. The LJB-sequent
$$[P(x) \imp P(y)]_{x,y}, [P(x)]_{x} \sequent P(z)$$ is then
translated as $$P(x') \imp P(y'), P(x'') \sequent P(z)$$
Notice that this LJB-sequent could also be translated as $$P(x_1)
\imp P(y_1), P(x_2) \sequent P(z)$$ Thus our translation will
not be a function mapping LJB-sequents to LJ$^{+}$-sequents, but
rather a relation between LJB-sequents and LJ$^{+}$-sequents.

In the rest of this section, we first define this relation, and then
prove the soundness and
completeness of LJB with respect to LJ$^{+}$.

\begin{definition}[Fresh $\alpha$-variants and flattening] 
Let $\Gamma \sequent A$ be a LJB-sequent, a {\em fresh
$\alpha$-variant} of $\Gamma \sequent A$ is a LJB-sequent
$\alpha$-equivalent to $\Gamma \sequent A$ satisfying Barendregt's
condition, {\em i.e.} where the bound variables are distinct and distinct
from the free variables.

A LJ$^{+}$-sequent $\Delta \sequent B$ is said to be a {\em
flattening} of a LJB-sequent $\Gamma \sequent A$, if it is
obtained by erasing all the brackets in a fresh $\alpha$-variant
of $\Gamma \sequent A$.
\end{definition}

\begin{definition}{\bf ($\overline{\alpha}$-equivalence)}
Two LJ$^{+}$-sequents $\Gamma \sequent A$ and $\Gamma'
\sequent A'$ are said to be {\em
${\overline{\alpha}}$-equivalent} if they differ only by the names
of some bound and free variables, {\em i.e.} if there exists two
substitutions $\sigma$ and $\sigma'$ such that $\sigma (\Gamma
\sequent A)$ is $\alpha$-equivalent to $\Gamma' \sequent A'$
and $\sigma' (\Gamma' \sequent A')$ is $\alpha$-equivalent to
$\Gamma \sequent A$.
\end{definition}

This relation is an equivalence relation. If two LJ$^{+}$-sequents
are flattenings of the same LJB-sequent, then they are
$\overline{\alpha}$-equivalent. 
For instance, the LJB-sequent
$$[P(x) \imp P(y)]_{x,y}, [P(x)]_{x} \sequent P(z)$$ 
has a flattening 
$$P(x') \imp P(y'), P(x'') \sequent P(z)$$
and also a flattening
$$P(x_1) \imp P(y_1), P(x_2) \sequent P(z)$$ 
and these two sequents are $\overline{\alpha}$-equivalent.

If two LJ$^{+}$-sequents are
$\overline{\alpha}$-equivalent then one has a derivation of height
$n$ if and only if the other does. Thus, if a flattening of a
LJB-sequent has a derivation of height $n$, then all do.

\begin{proposition}
\label{duplicate} If a LJ$^{+}$-sequent $\Gamma, A, A \sequent
B$ has a derivation in LJ$^{+}$, then so does $\Gamma, A
\sequent B$ and the derivations have the same height.
\end{proposition}

\proof{By induction on the structure of the derivation of $\Gamma,
A, A \sequent B$.}

\begin{proposition}
\label{invariant} Let $\Gamma$ and $\Gamma'$ be two LJB-contexts
such that $\Gamma \longrightarrow \Gamma'$ in the system of
Definition \ref{cleaning} and let $A$ be a formula. Let $\Delta
\sequent E$ be a flattening of $\Gamma \sequent A$ and
$\Delta' \sequent E'$ be a flattening of $\Gamma' \sequent
A$. Then, the LJ$^{+}$-sequent $\Delta \sequent E$ has a
derivation in LJ$^{+}$ if and only if the LJ$^{+}$-sequent
$\Delta' \sequent E'$ does and the derivations have the same
height.
\end{proposition}

\proof{For the first rule, the context $\Gamma$ has the form
$C([I, \Sigma]_{V})$ with $FV(I) \cap V = \emptyset$ and $\Gamma'
= C(I, [\Sigma]_{V})$. The LJ$^{+}$-sequent $\Delta \sequent E$
is obtained by erasing the brackets in a fresh $\alpha$-variant
$C'([I', \Sigma']_{V'}) \sequent E$ of $C([I, \Sigma]_{V})
\sequent A$. As $FV(I) \cap V = \emptyset$, the LJB-sequent
$C'(I',[\Sigma']_{V'}) \sequent E$ is a fresh $\alpha$-variant
of $C(I,[\Sigma]_{V}) \sequent A$ and thus $\Delta \sequent
E$ is also a flattening of $C'(I',[\Sigma']_{V'}) \sequent E$.
Thus $\Delta \sequent E$ and $\Delta' \sequent E'$ are two
flattenings of the same LJB-sequent. Hence they are
$\overline{\alpha}$-equivalent and if one has a derivation then so
does the other and the derivations have the same height.

The case of the second rule is trivial. For the third rule, the
context $\Gamma$ has the form $C(I,I)$ and $\Gamma' = C(I)$. The
LJ$^{+}$-sequent $\Delta \sequent E$ is obtained by erasing the
brackets in a fresh $\alpha$-variant $C'(I'_1,I'_2) \sequent E$
of $C(I,I) \sequent A$. Thus, $\Delta = \Sigma, \Xi_{1},
\Xi_{2}$ where $\Xi_1$ is the set obtained by erasing the brackets
in $I'_1$ and $\Xi_{2}$ in $I'_2$. Let $\overline{y}$ be the
variables bound by the brackets in $I'_1$. Then $\Xi_2$ has the
form $(\overline{y'}/\overline{y})\Xi_1$ for some variables
$\overline{y'}$. The sequent $C'(I'_2) \sequent E$ is a fresh
$\alpha$-variant of $C(I) \sequent A$, thus the sequent
$\Sigma, \Xi_{2} \sequent E$ is a flattening of $C(I)
\sequent A$. If $\Delta \sequent E$ has a derivation, then
substituting $\overline{y}$ by $\overline{y}'$ in this derivation
yields a derivation of the same height of the sequent $\Delta,
\Xi_2, \Xi_2 \sequent E$ and Proposition \ref{duplicate} yields
a derivation of the same height of $\Delta, \Xi_2 \sequent E$.
Thus, one flattening of $\Gamma' \sequent A$ has a derivation,
hence all do and the derivations have the same height. Conversely,
the sequent $\Delta' \sequent E'$ is obtained by erasing the
brackets in a fresh $\alpha$-variant $C'(I') \sequent E'$ of
$C(I) \sequent A$. Thus, $\Delta' = \Sigma', \Xi'$ where $\Xi'$
is the set obtained by erasing the brackets in $I'$. Let
$\overline{y}$ be the variables bound in $I'$ and $\overline{y}'$
be fresh variables. The sequent $\Sigma, \Xi',
(\overline{y}'/\overline{y})\Xi' \sequent E$ is a flattening of
$\Gamma \sequent A$. If $\Delta' \sequent E'$ has a
derivation, then so does $\Sigma, \Xi',
(\overline{y}'/\overline{y})\Xi' \sequent E$ and the
derivations have the same height. Thus, one flattening of $\Gamma
\sequent A$ has a derivation, hence all do and the derivations
have the same height.}

\begin{proposition}
\label{invariant21} Let $\Gamma$ and $\Delta$ be two LJB-contexts,
$A$ be a formula and $V$ be a set of variables such that $A$ has
no free variables in $V$. Let $\Sigma_1 \sequent E_1$ be a
flattening of $\Gamma, [\Delta]_V \sequent A$ and $\Sigma_2
\sequent E_2$ be a flattening of $[\Gamma]_V, [\Delta]_V
\sequent A$. Then the LJ$^{+}$-sequents $\Sigma_1 \sequent
E_1$ and $\Sigma_2 \sequent E_2$ are
$\overline{\alpha}$-equivalent.
\end{proposition}

\proof{As $A$ has no free variables in $V$, the LJB-sequent
$[\Gamma]_V, [\Delta]_V \sequent A$ has a fresh
$\alpha$-variant of the form $[\Gamma']_{V}, [(V'/V)\Delta']_{V'}
\sequent E'$, where the set $V$ is kept as subscript of the
brackets of $\Gamma'$. Let $\Sigma' \sequent E'$ be the
flattening of $[\Gamma]_V, [\Delta]_V \sequent A$ obtained by
erasing the brackets in the LJB-sequent $[\Gamma']_{V},
[(V'/V)\Delta']_{V'} \sequent E'$. The LJ$^{+}$-sequent
$\Sigma' \sequent E'$ is also obtained by erasing the brackets
in the LJB-sequent $\Gamma', [(V'/V)\Delta']_{V'} \sequent E'$
that is a fresh $\alpha$-variant of $\Gamma, [\Delta]_V
\sequent A$, thus it is also a flattening of the latter
LJB-sequent.

The LJ$^{+}$-sequents $\Sigma_1 \sequent E_1$ and $\Sigma'
\sequent E'$ are $\overline{\alpha}$-equivalent because they
are flattenings of the same LJB-sequent, and so are $\Sigma_2
\sequent E_2$ and $\Sigma' \sequent E'$. By transitivity,
the LJ$^{+}$-sequents $\Sigma_1 \sequent E_1$ and $\Sigma_2
\sequent E_2$ are $\overline{\alpha}$-equivalent.}

\begin{proposition} \label{invariant2}
Let $\Gamma$ and $\Delta$ be two LJB-contexts, $A$ be a formula
and $V$ be a set of variables such that $A$ has no free variables
in $V$. Let $\Sigma_1 \sequent E_1$ be a flattening of $\Gamma,
[\Delta]_V \sequent A$ and  $\Sigma_2 \sequent E_2$ a
flattening of $[\Gamma]_V, \Delta \sequent A$. Then, the
LJ$^{+}$-sequents $\Sigma_1 \sequent E_1$ and $\Sigma_2
\sequent E_2$ are $\overline{\alpha}$-equivalent.
\end{proposition}

\proof{As a corollary of Proposition \ref{invariant21}.}

\begin{proposition}{\bf (Soundness)}
\label{soundness} If the sequent $\sequent A$ has a derivation
in LJB, then it has also a derivation in LJ$^{+}$.
\end{proposition}

\proof{We prove, more generally, that if the LJB-sequent $\Gamma
\sequent A$ has a derivation in LJB then all its flattening
have a derivation in LJ$^{+}$. We proceed by induction on the
structure of the derivation of $\Gamma \sequent A$.

\begin{itemize}
\item If the last rule is $L\imp$
$$\irule{\Gamma' \sequent A_{1} ~~~ ...  ~~~\Gamma' \sequent A_{n}}
        {\Gamma \sequent A} {\mbox{$L\imp$}}$$
where
$$\Gamma = \Gamma_{1}, [\Gamma_{2}, [... \Gamma_{i-1},
[\Gamma_{i}, A_{1} \imp ... \imp A_{n} \imp A]_{V_{i-1}}
...]_{V_{2}}]_{V_{1}}$$
$$\Gamma' = ([...[[\Gamma_{1}]_{V_{1}}, \Gamma_{2}]_{V_{2}}, ...
\Gamma_{i-1}]_{V_{i-1}}, \Gamma_{i}, A_{1} \imp ... \imp A_{n} \imp
A) \downarrow $$ $A$ is atomic and has no free variables in $V_{1}
\cup V_{2} \cup ... \cup V_{i-1}$, then we consider a fresh
$\alpha$-variant of $\Gamma' \sequent A$. The variables bound
in this variant are not free in $A_{1}, ..., A_{n}$. Let $\Delta
\sequent E$ be the LJ$^{+}$-sequent obtained by erasing the
brackets in this variant. Let $E_1, ..., E_n$ be $\alpha$-variants
of $A_1, ..., A_n$ where the bound variables do not appear in
$\Gamma'$. The LJ$^{+}$-sequents $\Delta \sequent E_1$, ...,
$\Delta \sequent E_n$ are flattenings of $\Gamma' \sequent
A_1$, ..., $\Gamma' \sequent A_n$. Thus, by the induction
hypothesis, they have derivations in LJ$^{+}$. Applying the
$L\imp$ rule of LJ$^{+}$ we get a derivation of $\Delta
\sequent E$ and then of $\Delta \sequent A$ as $A$ and $E$
are $\alpha$-equivalent. Using Proposition \ref{invariant}, 
Proposition \ref{invariant2}, an induction on $i$ and the fact that 
$A$ has no free variables in $V_{1} \cup V_{2} \cup ... \cup V_{i-1}$,
we get a derivation of a flattening of $\Gamma \sequent A$. One
flattening of $\Gamma \sequent A$ is derivable, hence all are.

\item If the last rule is $R\forall$
$$\irule{[\Gamma]_{V}{\downarrow} \sequent B} {\Gamma \sequent \forall x~B}
{\mbox{$R\forall$}}$$ where $V$ is the set of all variables bound
in $\forall x~B$, then the variable $x$ is not free in
$[\Gamma]_{V}{\downarrow}$. Thus, the LJB-sequent
$[\Gamma]_{V}{\downarrow} \sequent B$ has a flattening $\Delta
\sequent B'$ such that the variable $x$ does not occur in
$\Delta$. By the induction hypothesis, $\Delta \sequent B'$ has
a derivation in LJ$^{+}$. As the variable $x$ does not occur free
in $\Delta$, we can apply the $R\forall$ rule of LJ$^{+}$ and
obtain a derivation of $\Delta \sequent \forall x~B'$. The
LJ$^{+}$-sequent $\Delta \sequent \forall x~B'$ is a flattening
of $[\Gamma]_{V}{\downarrow} \sequent \forall x~B$. By
Proposition \ref{invariant}, the LJB-sequent $[\Gamma]_{V}
\sequent \forall x~B$ has a derivable flattening.

Finally, notice that as $\forall x~B$ has no free variable in $V$,
every flattening of $[\Gamma]_{V} \sequent \forall x~B$ is
$\overline{\alpha}$-equivalent to a flattening of $\Gamma
\sequent \forall x~B$. Thus, the LJB-sequent $\Gamma
\sequent \forall x~B$ has a derivable flattening. One
flattening of $\Gamma \sequent \forall x~B$ is derivable, hence
all are.

\item If the last rule is $R\imp$
$$\irule{(\Gamma, B){\downarrow} \sequent C}
        {\Gamma \sequent B \imp C}
        {}$$
then, by the induction hypothesis, the LJB-sequent $(\Gamma,
B){\downarrow} \sequent C$ has a derivable flattening. By
Proposition \ref{invariant}, the LJB-sequent $\Gamma, B
\sequent C$ has a derivable flattening. This flattening has the
form $\Delta, B' \sequent C'$. Applying the $R\imp$ rule
of LJ$^{+}$, we get a proof of $\Delta \sequent B' \imp C'$ and
this LJ$^{+}$-sequent is a flattening of $\Gamma \sequent B \imp
C$. One flattening of $\Gamma \sequent B \imp C$ is derivable,
hence all are.
\end{itemize}
}

\begin{proposition}{\bf (Completeness)}
\label{completeness} If the sequent $\sequent A$ has a
derivation in LJ$^{+}$, then it also has a derivation in LJB.
\end{proposition}

\proof{We prove, more generally, that if a flattening $\Delta
\sequent E$ of $\Gamma \sequent A$ has a derivation $\pi$ in
LJ$^{+}$, then the LJB-sequent $\Gamma \sequent A$ has a
derivation in LJB. We proceed by induction on the height of the
derivation $\pi$.

\begin{itemize}

\item If the last rule is $L\imp$ then $E$ is atomic, the
formul\ae\  $A$ and $E$ are identical, $\Delta$ contains a formula
of the form $A_{1} \imp ... \imp A_{n} \imp A$ and the sequents
$\Delta \sequent A_{1}, ..., \Delta \sequent A_{n}$ have
derivations smaller than $\pi$.

Thus, the context $\Gamma$ contains a formula $B$, corresponding
to the formula $A_{1} \imp ... \imp A_{n} \imp A$ through flattening,
and $\Gamma$ has the form $\Gamma = \Gamma_{1}, [\Gamma_{2}, [...
\Gamma_{i-1}, [\Gamma_{i}, B]_{V_{i-1}} ...]_{V_{2}}]_{V_{1}}$.

The formula $B$ has the form $C_{1} \imp ... \imp C_{n} \imp C$,
where $C$ is an atomic formula. Renaming in $C$ the variables of
$V_{1} \cup V_{2} \cup ... \cup V_{i-1}$ with variables not free
in $A$ yields the formula $A$. Hence $C = A$ and has no free
variables in $V_{1} \cup V_{2} \cup ... \cup V_{i-1}$.

Let $\Gamma^* = [...[[\Gamma_{1}]_{V_{1}}, \Gamma_{2}]_{V_{2}},
..., \Gamma_{i-1}]_{V_{i-1}}, \Gamma_{i}, B$ and $\Delta'
\sequent A$ be a flattening of $\Gamma^* \sequent A$.
Using Proposition \ref{invariant2}, an induction on $i$ and the fact
that $A$ has no free variables in $V_{1} \cup V_{2} \cup ... \cup
V_{i-1}$, we get that
the
LJ$^{+}$-sequents $\Delta \sequent A$ and $\Delta' \sequent
A$ are $\overline{\alpha}$-equivalent. Thus, there exists a
substitution $\sigma$ such that $\sigma (\Delta \sequent A)$ is
$\alpha$-equivalent to $\Delta' \sequent A$. The formula
$\sigma A_{i}$ is $\alpha$-equivalent to $C_{i}$.

The LJ$^{+}$-sequents $\Delta \sequent A_{i}$ have derivations
smaller than $\pi$ thus, so do the LJ$^{+}$-sequents
$\sigma(\Delta \sequent A_{i})$, {\em i.e.} $\sigma \Delta
\sequent C_{i}$. Let $C'_i$ be  an $\alpha$-variant of $C_i$
where the bound variables do not appear in $\sigma \Delta$. The
sequents $\sigma \Delta \sequent C'_i$ have derivations smaller
than $\pi$ and they are flattenings of $\Gamma^* \sequent C_i$.

Thus, the LJB-sequents $\Gamma^{*} \sequent C_{i}$ have
flattenings that have derivations smaller than $\pi$. By
Proposition \ref{invariant}, so do the sequents
$\Gamma^*{\downarrow} \sequent C_{1}$, ...,
$\Gamma^*{\downarrow} \sequent C_{n}$. By the induction
hypothesis, the sequents $\Gamma^*{\downarrow} \sequent C_{1}$,
..., $\Gamma^*{\downarrow} \sequent C_{n}$ are derivable in LJB
and we conclude with the $L\imp$ rule of LJB.

\item If the last rule is $R\forall$, then the formula $E$ has the
form $\forall x~B$, the variable $x$ does not occur free in
$\Delta$ and the LJ$^{+}$-sequent $\Delta \sequent B$ has a
derivation smaller than $\pi$. The formula $A$ is
$\alpha$-equivalent to $E$ and has the form $\forall y~B'$.

Let $V$ be the set of variables bound in $A$. As stated in the
definition of LJB, the free and bound variables of $A$ are
disjoint and the free variables of $A$ and $E$ are the same. Thus
the bound variables of $A$ are not free in $E = \forall x~B$, and
$V - \{x\}$ and $FV(B)$ are disjoint.

Let $\sigma$ be a substitution renaming all the variables of $V$
with fresh variables and $\sigma'$ its restriction to $V - \{x\}$.

As the LJ$^{+}$-sequent $\Delta \sequent B$ has a derivation
smaller than $\pi$, so does the LJ$^{+}$-sequent $\sigma' \Delta
\sequent \sigma' B$.

As the domain of $\sigma'$ and $FV(B)$ are disjoint, $\sigma' B =
B$. Moreover as $x$ is not free in $\Delta$, we have $\sigma
\Delta = \sigma' \Delta$. Thus, the LJ$^{+}$-sequent $\sigma
\Delta \sequent B$ has a derivation smaller than $\pi$.

The LJ$^{+}$-sequent $\sigma \Delta \sequent B$ is
$\overline{\alpha}$-equivalent to a flattening of $[\Gamma]_V
\sequent B'$. Thus, the sequent $[\Gamma]_V \sequent B'$ has
a flattening that has a derivation smaller than $\pi$. By
Proposition \ref{invariant}, so does the sequent $[\Gamma]_V
{\downarrow} \sequent B'$. By the induction hypothesis, the
sequent $[\Gamma]_V {\downarrow} \sequent B'$ has a derivation
in LJB and we conclude with the $R\forall$ rule of LJB.

\item If the last rule is $R\imp$ then the formula $E$ has
the form $B \imp C$, the formula $A$ has the form $B' \imp C'$ where
$B$ is $\alpha$-equivalent to $B'$ and $C$ to $C'$, and the
LJ$^{+}$-sequent $\Delta, B \sequent C$ has a derivation
smaller than $\pi$. The LJ$^{+}$-sequent $\Delta, B \sequent C$
is a flattening of $\Gamma, B' \sequent C'$. Thus, the
LJB-sequent $\Gamma, B' \sequent C'$ has a flattening that has
a derivation smaller than $\pi$. By Proposition \ref{invariant},
the LJB-sequent $(\Gamma, B'){\downarrow} \sequent C'$ has a
flattening that has a derivation smaller than $\pi$. By the
induction hypothesis, the LJB-sequent $(\Gamma, B'){\downarrow}
\sequent C'$ has a derivation in LJB and we conclude with the
$R\imp$ rule of LJB.
\end{itemize}
}

\section{Decidability}

To show that provability in the system LJB is decidable, we
consider a closed formula $E$ where all bound variables are
distinct. The formul\ae\  occurring in a derivation of
$\sequent E$ in LJB are pieces of $E$.

A position $f$ of $E$ is said to be {\em in the scope} of a
variable $x$ if the unique position of $E$ labeled by $\forall
x$ is a strict prefix of $f$.

A variable $y$ is said to be {\em in the scope} of $x$ if the
unique position of $E$ labeled by $\forall x$ is a strict
prefix of the unique position of $E$ labeled by $\forall y$.
This relation is obviously transitive.

Let $V(x)$ be the set of all variables bound in the unique piece
of $E$ of the form $\forall x~A$. This set can alternatively be
defined as the set containing $x$ and the variables $y$'s in the
scope of $x$ in $E$. All the sets of variables occurring as a
subscript of brackets in a derivation of $\sequent E$ have the
form $V(x)$ for some variable $x$ of $E$.

As $E$ is a closed formula where all the bound variables
are distinct, if a variable $x$ occurs free in the formula
associated to a position $f$ of $E$, then $f$ is in the scope of
$x$.

\begin{proposition}
\label{tricot} If the position $f$ is in the scope of a variable
$x$ and the formula associated to $f$ has a free occurrence of a
variable $y$ then either $x = y$ or $x$ is in the scope of $y$ or
else $y$ is in the scope of $x$.
\end{proposition}

\proof{Let $g$ be the unique position of $E$ labeled by $\forall
x$ and $h$ the unique position of $E$ labeled by $\forall y$.
Both $g$ and $h$ are prefixes of $f$. Hence either $g = h$ or $h$
is a strict prefix of $g$ or else $g$ is a strict prefix of $h$.}

\begin{proposition}
\label{tricot2} If the variable $z$ is in the scope both of $x$
and of $y$ then either $x = y$ or $x$ is in the scope of $y$ or
else $y$ is in the scope of $x$.
\end{proposition}

\proof{Let $f$ be the unique position of $E$ labeled by $\forall
x$, $g$ the unique position of $E$ labeled by $\forall y$ and $h$
the unique position of $E$ labeled by $\forall z$. Both $f$ and
$g$ are prefixes of $h$. Hence either $f = g$ or $g$ is a strict
prefix of $f$ or else $f$ is a strict prefix of $g$.}

\begin{definition}{\bf (Depth of a LJB-context and of an item)}
The {\em depth} of a LJB-context is defined by
\begin{itemize}
\item $depth(\{I_{1}, ..., I_{n}\}) =  max \{depth(I_{1}) , ...,
 depth(I_{n}) \}$,
\end{itemize}
and the {\em depth} of an item is defined by
\begin{itemize}
\item $depth(A) =  0$, \item $depth([\Gamma]_{V}) = 1 + depth(\Gamma)$.
\end{itemize}
\end{definition}

\begin{proposition}
\label{scopefv} Let $[\Gamma]_{V(x)}$ be a normal item occurring
in a derivation of $\sequent E$ in LJB and $z$ a free variable of
$[\Gamma]_{V(x)}$. Then $x$ is in the scope of $z$.
\end{proposition}

\proof{By induction on the depth of $[\Gamma]_{V(x)}$. First, note
that the variable $z$ occurs free in an item $I$ of $\Gamma$ and
is not a member of $V(x)$.
As $[\Gamma]_{V(x)}$ is normal, $I$ has a free variable $y$ in the
set $V(x)$.

If $I$ is a formula then let $f$ be its occurrence in $E$. The
occurrence $f$ is in the scope of $y$ in $E$. As $y$
is in $V(x)$ then either $y = x$ or $y$ is in the scope of $x$. In
both cases, $f$ is in the scope of $x$ in $E$. Hence, by
Proposition \ref{tricot}, as the variable $z$ is free in $I$,
either $x = z$ or $x$ is in the scope of $z$ or $z$ is in the
scope of $x$. As, moreover, $z$ is not in $V(x)$ then $x$ is in
the scope of $z$.

If $I$ is itself an item of the form $[\Gamma']_{V(x')}$, then, by
the induction hypothesis $x'$ is in the scope of all the free
variables of $[\Gamma']_{V(x')}$, and in particular $x'$ is in the
scope of $y$ and $z$. As $y$ is in $V(x)$ then either $y = x$ or
$y$ is in the scope of $x$. In both cases $x'$ is in the scope of
$x$. Hence by Proposition \ref{tricot2}, either $x = z$ or $x$ is
in the scope of $z$ or $z$ is in the scope of $x$. As, moreover,
$z$ is not in $V(x)$ then $x$ is in the scope of $z$.}

\begin{proposition}
\label{scope2}
 Let $[\Gamma]_{V(x)}$ be a normal item. For every
item of $\Gamma$ of the form $[\Gamma']_{V(x')}$, the variable
$x'$ is in the scope of $x$.
\end{proposition}

\proof{As $[\Gamma]_{V(x)}$ is normal, $[\Gamma']_{V(x')}$ has a
free variable $y$ in the set $V(x)$ and, by Proposition
\ref{scopefv}, $x'$ is in the scope of $y$. As $y$ is in $V(x)$
then either $y = x$ or $y$ is in the scope of $x$. In both cases
$x'$ is in the scope of $x$.}

\begin{proposition}
\label{depth} Let $E$ be a closed formula, $S$ the finite set of
the pieces of $E$ and $d$ the maximum depth of nested variables in
$E$. Let $T$ be the finite set of LJB-sequents formed with
formul\ae\ of $S$, whose subscripts are of the form $V(x)$ for
some variable $x$ of $E$ and whose depth is bounded by $d$. Then,
only LJB-sequents of $T$ can occur in a proof of $\sequent E$.
\end{proposition}

\proof{As already noticed, all the formul\ae\ occurring in a
derivation of $\sequent E$ are pieces of $E$ and all subscripts
occurring in a derivation of $\sequent E$ are of the form
$V(x)$ for some variable $x$ of $E$. By Proposition \ref{scope2},
the depth of the LJB-sequents occurring in a derivation of
$\sequent E$ is bounded by $d$. }

\begin{proposition}
\label{nonredundant} If a LJB-sequent $\Gamma \sequent A$ has a
derivation, then it has a non redundant derivation, {\em i.e.} a
derivation where the same sequent does not occur twice in the same
branch.
\end{proposition}

\proof{By induction on the number of sequents occurrences
in the proof. Consider a redundant proof where the LJB-sequent $\Gamma
\sequent A$ occurs twice in the same branch. We can replace the bigger
proof of this sequent by the smaller one, yielding a smaller proof, to
which we apply the induction hypothesis.}

\medskip

The following proposition is a straightforward consequence of
Propositions \ref{depth} and \ref{nonredundant}.

\begin{proposition}
Provability in the system LJB is decidable.
\end{proposition}

\begin{remark}
If $n$ is the size of the formula $A$, then the cardinal of
 $S$ is exponential in $n$ and that of $T$
doubly exponential, where $S$ and $T$ are as defined in
Proposition \ref{depth}. Thus a doubly exponential decision
algorithm can be obtained from any algorithm visiting each sequent
at most once.
\end{remark}

\medskip

\begin{remark}
This decidability proof uses the fact that the $R\forall$ rule
binds all the variables bound in the right hand side of the
sequent and not just $x$. The termination of proof search in the
simpler system where this rule binds the variable $x$ only is left
open.
\end{remark}

\section{Application to simple type theory and system F}

In \cite{DowekHardinKirchner} we have given a presentation of
simple type theory (higher-order logic) as a theory in first-order
predicate logic. We have also given a presentation of this theory
in deduction modulo \cite{TPM} where axioms are replaced by
rewrite rules. For instance when we have a formula $\forall
x~\varepsilon(x)$ and we substitute $x$ by the term
$\dot{\imp}(y,z)$ we have to normalize the formula
$\varepsilon(\dot{\imp}(y,z))$ yielding $\varepsilon(y) \imp
\varepsilon(z)$. We have shown that simple type theory can be
presented with rewrite rules only and no axioms.

When we have a theory in deduction modulo formed by a confluent
and terminating rewrite system and no axioms and with the cut
elimination property, we can decide if a positive normal formula
is provable or not in this theory. Indeed, as we never substitute
variables in a derivation, normal formul\ae\  remain normal and
the rewrite rules can never be used. Thus, a normal formula is
provable in this theory if and only if it is provable in predicate
logic.

Thus, inhabitation in the positive minimal fragment of simple type
theory is decidable.

We obtain also this way a new decidability proof for the positive
fragment of system F \cite{Jiang}, while the general inhabitation
problem for system F is known to be undecidable \cite{Loeb}.

\begin{proposition}
Inhabitation in the positive fragment of system F is decidable.
\end{proposition}

\proof{To each type of system $F$ we associate a formula in
minimal predicate logic, with a single unary predicate
$\varepsilon$ as in \cite{DowekHardinKirchner}.
$$\Phi(X) = \varepsilon(X)$$
$$\Phi(T \imp U) = \Phi (T) \imp \Phi (U)$$
$$\Phi(\forall X~T) = \forall X~\Phi(T)$$

For instance $\Phi (\forall X~(X \imp X)) = \forall
X~(\varepsilon(X) \imp \varepsilon(X))$.

As there is no substitution of variables in the positive fragment,
a positive type $T$ is inhabited in system $F$ if and only if the
formula $\Phi(T)$ is provable in minimal predicate logic. Thus
inhabitation for positive types in system $F$ is decidable. }

Let us consider some examples. The system LJB allows to show that
the type of Example 1 is empty in System F, while that of Example
2, its prenex form, is inhabited. For ease of reading, we write
\(X\) instead of \(\varepsilon(X)\).

\noindent {\it Example 1.} Let us try to prove the inhabitation of the type

\[
\forall X~(((\forall Y \forall Z~(((Y \imp X) \imp Z)\imp (Y \imp Z)
\imp Z)) \imp X)\imp X)
\]

\noindent Let $C(X) = (\forall Y\forall Z~(((Y \imp X) \imp Z) \imp (Y \imp
Z) \imp Z)) \imp X$.

{\footnotesize
$$\irule{
\irule{\irule{\irule{\irule{\irule{\irule{\irule{\irule{\irule{\irule{\irule{...}
{C(X), [(Y \imp X) \imp Z, Y \imp Z, Y]_{YZ}, (Y \imp X) \imp Z, Y \imp Z
\sequent Z} {} } {C(X), [(Y \imp X) \imp Z, Y \imp Z, Y]_{YZ}, (Y \imp
X) \imp Z, Y \imp Z, Y \sequent \forall Y\forall Z (((Y \imp X) \imp Z)
\imp (Y \imp Z) \imp Z)} {\mbox{$R\imp$, $R\forall$}} } {C(X), [(Y
\imp X) \imp Z, Y \imp Z, Y]_{YZ}, (Y \imp X) \imp Z, Y \imp Z, Y
\sequent X} {\mbox{$L\imp$}} } {C(X), [(Y \imp X) \imp Z, Y
\imp Z, Y]_{YZ}, (Y \imp X) \imp Z, Y \imp Z \sequent Y \imp X}
{\mbox{$R\imp$}} } {C(X), [(Y \imp X) \imp Z, Y \imp Z, Y]_{YZ},
(Y \imp X) \imp Z, Y \imp Z \sequent Z} {\mbox{$L\imp$}} }
{C(X), (Y \imp X) \imp Z, Y \imp Z, Y \sequent \forall Y\forall Z (((Y
\imp X) \imp Z) \imp (Y \imp Z) \imp Z)} {\mbox{$R\imp$,
$R\forall$}} } {C(X), (Y \imp X) \imp Z, Y \imp Z, Y \sequent X}
{\mbox{$L\imp$}} } {C(X), (Y \imp X) \imp Z, Y \imp Z \sequent
Y \imp X} {\mbox{$R\imp$}} } {C(X), (Y \imp X) \imp Z, Y \imp Z
\sequent Z} {\mbox{$L\imp$}} } {C(X) \sequent \forall
Y\forall Z (((Y \imp X) \imp Z) \imp (Y \imp Z) \imp Z)} {\mbox{$R\forall$
$R\imp$}} } {C(X) \sequent X} {\mbox{$L\imp$}} }
{\sequent \forall X (C(X) \imp X)} {\mbox{$R\imp$,
$R\forall$}}$$} Again, restricting to non redundant proofs prunes this
branch. We can check that the other branches are pruned in the same
way.  Thus, this type is empty.\\

\noindent {\it Example 2.} In contrast trying to prove the inhabitation of
the type
\[\forall X \forall Y \forall Z~(((((Y \imp X) \imp Z) \imp (Y \imp Z) \imp Z) \imp X)\imp X)
\]
yields the following derivation
$$\irule{\irule{
\irule{\irule{\irule{\irule{\irule{\irule{\irule{\irule{}
                                    {(((Y \imp X) \imp Z) \imp (Y \imp Z) \imp Z) \imp X, (Y \imp X) \imp Z, Y \imp Z, Y \sequent Y}                                                                            {\mbox{$L\imp$}}
                                                             }
                               {(((Y \imp X) \imp Z) \imp (Y \imp Z) \imp Z) \imp X, (Y \imp X) \imp Z, Y \imp Z, Y \sequent Z}
                                                         {\mbox{$L\imp$}}
                                                  }
                                     {(((Y \imp X) \imp Z) \imp (Y \imp Z) \imp Z) \imp X, (Y \imp X) \imp Z, Y \imp Z, Y \sequent ((Y \imp X) \imp Z) \imp (Y \imp Z) \imp Z}
                                                  {\mbox{$R\imp$}}
                                           }
                                           {(((Y \imp X) \imp Z) \imp (Y \imp Z) \imp Z) \imp X, (Y \imp X) \imp Z, Y \imp Z, Y \sequent X}
                                           {\mbox{$L\imp$}}
                                    }
                      {(((Y \imp X) \imp Z) \imp (Y \imp Z) \imp Z) \imp X, (Y \imp X) \imp Z, Y \imp Z \sequent Y \imp X}
        {\mbox{$R\imp$}}
               }
               {(((Y \imp X) \imp Z) \imp (Y \imp Z) \imp Z) \imp X, (Y \imp X) \imp Z, Y \imp Z \sequent Z}
               {\mbox{$L\imp$}}
        }
        {(((Y \imp X) \imp Z)\imp (Y \imp Z) \imp Z) \imp X \sequent ((Y \imp X) \imp Z) \imp (Y \imp Z) \imp Z}
        {\mbox{$R\imp$}}
  }
  {(((Y \imp X) \imp Z) \imp (Y \imp Z) \imp Z) \imp X \sequent X}
  {\mbox{$L\imp$}}
} {\sequent (((Y \imp X) \imp Z) \imp (Y \imp Z) \imp Z) \imp X \imp
X} {\mbox{$R\imp$}} } {\sequent \forall X \forall Y
\forall Z~((((Y \imp X) \imp Z) \imp (Y \imp Z) \imp Z) \imp X \imp X)}
{\mbox{$R\forall$ (3)}}$$
\bigskip

\noindent Thus, this type is inhabited.

\section{An implementation}

\begin{figure}
\noindent {\footnotesize
\begin{verbatim}
type term = |Var of string
            | Func of string * term list;;
type prop = | Atomic of string * term list
            | Imp of prop * prop
            | Forall of string * prop;;
type item = | Prop of prop
            | Bracket of item list * string list;;

let rec bv (p:prop) = match p with
| Atomic(_,l) -> []
| Imp(p1,p2)  -> (bv p1)@(bv p2)
| Forall(x,p) -> x::(bv p);;

let rec disjt (v:string list) (t:term) = match t with
| Var x -> not (List.mem x v)
| Func(_,l) -> List.for_all (disjt v) l;;
let rec disjp (v:string list) (p:prop) = match p with
| Atomic(_,l) -> List.for_all (disjt v) l
| Imp(p1,p2)  -> (disjp v p1) && (disjp v p2)
| Forall(x,p) -> disjp (List.filter (fun y -> not (x = y)) v) p;;
let rec disji (v:string list) (i:item) = match i with
| Prop p          -> disjp v p
| Bracket(l,vars) ->
      List.for_all (disji (List.filter (fun y -> not (List.mem y vars)) v)) l;;

let rec decompose (p:prop) = match p with
| Atomic(s,l) -> p,[]
| Imp(a1,a2)  -> let (h,t) = decompose a2 in (h,a1::t)
| Forall _ -> failwith "negative";;

let rec red (l:item list) = match l with
| a::b::l' -> if (a = b) then red (a::l') else a::(red (b::l'))
| _        -> l;;

let fuse (l1:item list) (l2:item list) = red (List.merge compare l1 l2);;

let rec bracket (l:item list) (v:string list) =
 let (l1,l2) = out v l
 in if l1 = [] then l2 else fuse [Bracket(l1,v)] l2
and out (v:string list) (l:item list) = match l with
| []    -> [],[]
| a::l' -> let (l1,l2) = out v l'
           in if disji v a then (l1,fuse [a] l2) else (fuse [a] l1,l2);;

let rec der (seen:(item list * prop) list) (g:item list) (p:prop) =
not (List.mem (g,p) seen) &&
let seen' = (g,p)::seen
in match p with
| Atomic(s,l)  -> some seen' g [] p
| Imp(a,b)     -> der seen' (fuse [Prop(a)] g) b
| Forall (x,a) -> der seen' (bracket g (bv p)) a
and some (seen:(item list * prop) list) (g:item list) (g1:item list)
(p:prop) = match g with
| []            -> false
| (Prop(p'))::g' ->
  let (h,t) = decompose p'
  in ((h = p) && (List.for_all (der seen (fuse g g1)) t))
     || (some seen g' (fuse [Prop(p')] g1) p)
| (Bracket(l1,v))::g' ->
  ((disjp v p) && (some seen l1 (bracket (fuse g g1) v) p))
  || (some seen g' (fuse [Bracket(l1,v)] g1) p)
and derivable (p:prop) = der [] [] p;;
\end{verbatim}
} \caption{An implementation} \label{implementation}
\end{figure}

A naive implementation in Objective Caml, version 3.08, is given
in Figure \ref{implementation}. Notice that in this program we do
not visit each sequent at most once, but merely search for a non
redundant proof. We do not use an explicit cleaning function, but
rather maintain all contexts clean with the help of two functions:
the function {\tt fuse}, that from two clean contexts {\tt l1} and
{\tt l2} builds the normal form of the context {\tt l1} $\cup$
{\tt l2}, and the function {\tt bracket} that from a clean context
{\tt l} and a set of variables {\tt v} builds the normal form of
the context {\tt [l]$_{\tt v}$}.

Using this implementation, we can, for example, check that the
formula
$$((\forall x~(P(x) \imp ((\forall y~(P(y) \imp Q))
\imp R) \imp R)) \imp Q) \imp Q$$ is not derivable.

\begin{verbatim}
derivable
  (Imp(Imp(Forall("x",Imp(Atomic("P",[Var("x")]),
                          Imp(Imp (Forall ("y",Imp (Atomic("P",[Var("y")]),
                                                    Atomic("Q",[]))),
                                    Atomic("R",[])),
                               Atomic("R",[])))),
           Atomic("Q",[])),
      Atomic ("Q",[])));;
\end{verbatim}

\begin{verbatim}
- : bool = false
\end{verbatim}

\section*{Conclusion}

It is well known that variable names are irrelevant in logic and
that they can be replaced by other scoping mechanisms. We have
shown in this paper that replacing the eigenvariable condition by
an appropriate bracketing mechanism simplifies the decision
algorithm of the positive part of minimal predicate logic.

This bracketing mechanism could be used in other situations where
we need to handle fresh variables, but its generality still needs
to be investigated.

\section*{Acknowledgments}

The authors want to thank the anonymous referees for very helpful
suggestions that helped to improve the paper a lot. This work is
partially supported by NSFC 60373050, NSFC 60421001 and NSFC
60310213.

\end{document}